# A method of non-linear state feedback controller design based on state prediction

Paek Su-Yong, Ri Jin-Song

**Kim Il Sung** University, Pyongyang, DPR of Korea

**Abstracts**

In this paper, we considered a design method of non-linear state feedback controller for input-affine non-linear system taking data losses into account.

When data is lost in control system, control input is fixed to constant value or to the last implemented value [1,2].

When data losses are taken into account, these methods don't guarantee the stability of desired closed-loop.

The proposed method guarantees the stability and robustness properties in the presence of uncertainty and data losses.

**Keywords**: networked control system, state prediction, state feedback controller.

## 1. Introduction

Most control systems are designed under the assumption, of flawless communication at the sensor-controller and controller-actuator links and continuous or synchronous measurement sampling. These assumptions hold in most applications where point-to-point communication links are used and measurements of velocity, position or temperature are fed into the control system.

Control systems which operate over a communication network (wired or wireless) are called networked control systems (NCS) and can substantially improve the efficiency, flexibility, robustness and fault-tolerance of an industrial control system as well as reduction of the installation, reconfiguration and maintenance costs.

In general, network dynamics are modeled as one with time-varying delays, data quantization or data losses.

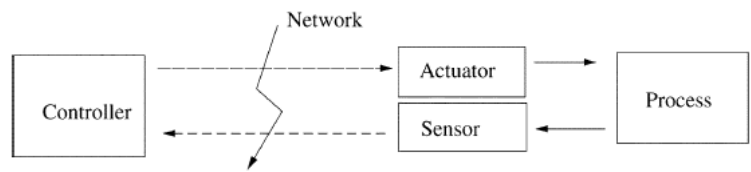

Fig1. Networked control system subject to data losses.

Fig. 1 shows a schematic diagram of a NCS, where data can be lost at the sensor-controller and controller-actuator links. In both cases feedback is lost, and the actuator must operate on its own, usually setting the control input to zero or to the last implemented value.

In [5], stability and disturbance attenuation issues for a class of linear NCS subject to data losses modeled as a discrete-time switched linear system with arbitrary switching was studied. In [3] (see also [7]–[9]) optimal control of linear time-invariant systems over unreliable communication links under different communication protocols (with and without acknowledgement of successful communication) was investigated and sufficient conditions for the existence of stabilizing control laws were derived.

In [4], the stability properties of a class of NCS modeled as linear asynchronous systems were studied.

In [6], general nonlinear NCS with disturbances are studied.

This paper is organized as follows. In Section II the class of nonlinear systems considered in this work is introduced and proposed state predictive method based on plant model. In Section III, we considered state feedback controller design method with data losses. In Section IV, the theoretical results are demonstrated through a pressure control of chemical reactor example.

## 2. State prediction based on plant model.

We consider SISO input affine non-linear system as follows.

$$\dot{x} = f(x) + g(x)u \qquad (1)$$
$$x(t_0) = x_0$$

We introduce time step $\Delta$ and break point $t_i = t_0 + i\Delta$, and calculate increments $\Delta x_i$ at break point $t_i$ as following equation.

$$\Delta x_i = \frac{1}{6}(k_1(i) + 2k_2(i) + 2k_3(i) + k_4(i)) \qquad (2)$$

Here, $k_1(i), k_2(i), k_3(i), k_4(i)$ is calculated as follows.

$$k_1(i) = \Delta * [f(x_i) + g(x_i) * u(i)] \qquad (3)$$

$$k_2(i) = \Delta * \left[ f\left(x_i + \frac{1}{2}k_1(i)\right) + g\left(x_i + \frac{1}{2}k_1(i)\right) * u(i) \right] \qquad (4)$$

$$k_3(i) = \Delta * \left[ f\left(x_i + \frac{1}{2}k_2(i)\right) + g\left(x_i + \frac{1}{2}k_2(i)\right) * u(i) \right] \qquad (5)$$

$$k_4(i) = \Delta * [f(x_i + k_3(i)) + g(x_i + k_3(i)) * u(i)] \qquad (6)$$

Then state predictive value at break point $t_{k+1}$ is calculated as follows.

$$\hat{x}_{i+1} = \hat{x}_i + \Delta x_i, \; x(t_0) = x_0 = \hat{x}_0 \qquad (7)$$

$\Delta$ is selected to guarantee convergence of algorithm.

In order to reduce predictive error, introduce new variable $\gamma(|\gamma| < 1)$.

State prediction is calculated as follows.

$$\hat{x}_{i+1} = (1 + \gamma)\hat{x}_i + \Delta x_i \qquad (8)$$

Here, value of $\gamma$ is chosen suitably in accordance with plant.

Two method of choosing $\gamma$ is as follows.

**[Method one]**

First, gets predictive value samples and actual measurement value samples based on nominal model of plant.

Let the number of sample to be N.

Let predictive value samples $\{\hat{x}\}$, actual measurement value samples $\{x\}$.

Average squared error E is calculated as

$$E = \frac{1}{N}\sum_{i=1}^{N}(x_i - \hat{x}_i)^2 \qquad (9)$$

And $\gamma$ as

$$\gamma = \zeta \frac{E}{\frac{1}{N}\sum_{i=1}^{N}\hat{x}_i} \qquad (10)$$

$\zeta$ is as below

$$\zeta = \begin{cases} 1 & \frac{1}{N}\sum_{i=1}^{N}\hat{x}_i \le \frac{1}{N}\sum_{i=1}^{N}x_i \\ -1, & \frac{1}{N}\sum_{i=1}^{N}\hat{x}_i > \frac{1}{N}\sum_{i=1}^{N}x_i \end{cases} \quad (11)$$

**[Method two]**

Gets predictive value samples and actual measurement value samples based on nominal model of plant.

Let the number of sample to be N.

Let predictive value samples $\{\hat{x}\}$, actual measurement value samples $\{x\}$.

Let $\{\{\hat{x}\},\{x\}\}$ to be samples pair.

$$\{\{\hat{x}_1\},\{x_1\}\},\{\{\hat{x}_2\},\{x_2\}\},\wedge\{\{\hat{x}_m\},\{x_m\}\} \quad (12)$$

Here $\{\{\hat{x}_i\},\{x_i\}\}, i=1,2,\wedge,m$ means $i-th$ samples pair and $\{\hat{x}_i\},\{x_i\}$ means $i-th$ predictive value samples and real value samples respectively.

Average squared error $E_i (i=1,2,...,m)$ is calculated as

$$E_i = \frac{1}{N}\sum_{k=1}^{N}(x_{ik}-\hat{x}_{ik})^2 \quad (13)$$

Average error $E$ is calculated as

$$E = \frac{1}{m}\sum_{i=1}^{m}E_i \quad (14)$$

And $\gamma$ as

$$\gamma = \zeta \frac{E}{\frac{1}{m}\sum_{i=1}^{m}\left(\frac{1}{N}\sum_{k=1}^{N}\hat{x}_{ik}\right)} \quad (15)$$

$$\zeta = \begin{cases} 1 & \frac{1}{m}\sum_{i=1}^{m}\left(\frac{1}{N}\sum_{k=1}^{N}\hat{x}_{ik}\right) \le \frac{1}{m}\sum_{i=1}^{m}\left(\frac{1}{N}\sum_{k=1}^{N}x_{ik}\right) \\ -1 & \frac{1}{m}\sum_{i=1}^{m}\left(\frac{1}{N}\sum_{k=1}^{N}\hat{x}_{ik}\right) > \frac{1}{m}\sum_{i=1}^{m}\left(\frac{1}{N}\sum_{k=1}^{N}x_{ik}\right) \end{cases} \quad (15\text{-}1)$$

## 3. State feedback controller design with data losses.

Consider SISO input-affine nonlinear system.

$$\dot{x} = f(x) + g(x)u + w(x)\theta \quad (16)$$

Here, $f(x), g(x), w(x)$ is $C^{\infty}$-function and $w(x)\theta$ represents uncertainty.

The following feedback law [formula of SonTag] asymptotically stabilizes the open-loop unstable steady-state of the nominal system[2].

$$h(x) = \begin{cases} -\dfrac{L_f V + \sqrt{(L_f V)^2 + \|L_g V\|^4}}{\|L_g V\|^2}, & L_g V \neq 0 \\ 0, & L_g V = 0 \end{cases} \quad (17)$$

Where $L_f V$ and $L_g V$ denote the Lie derivatives of the scalar function $V$ with respect to the vectors fields $f$ and $g$, respectively.

$$L_g V = \frac{\partial V}{\partial x} g(x), \quad L_f V = \frac{\partial V}{\partial x} f(x) \quad (18)$$

In this paper, we design predictive controller with data losses based on above state feedback controller. Fig2 shows control system diagram.

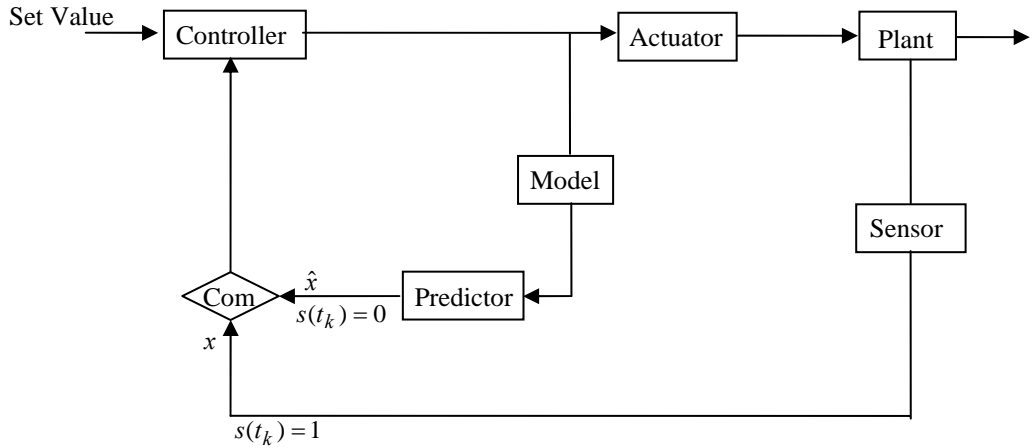

Fig2. control system diagram

Non-linear state feedback controller design procedure is summarized in the following algorithm.

**[Non-linear state feedback controller algorithm with data losses]**

Step1: design the state feedback controller of nonlinear system. Equation(17)

Step2: design state predictor based on the plant model Equation (8)

Step3: calculate control input trajectory at sample instant $t_k$ using the designed controller (17) and state predictor (8). $i = 0$

Step4: if $s(t_k) = 1$ then $u_k(t) = u_k(t_k), t \in [t_k, t_{k+1}], i = 0$

if $s(t_k) = 0$ then $u_k(t) = u_{k-i-1}(t_{k+i}), t \in [t_k, t_{k+1}], i = i+1$

if $i > N$ then $i = N$

Step5: $u(t) = u_k(t), t \in [t_k, t_{k+1}]$

Step6: obtain a new sample( $k = k+1$ ) and go to step4

## 4. Simulation and result analysis.

We verified the effectiveness of proposed method by applying to pressure control of chemical reactor tank.

Fig3 shows device configuration of tank pressure control.

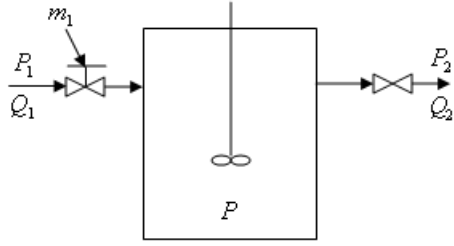

Fig3. configuration of tank pressure control

Mathematical model of pressure control plant is as follows.

$$C\frac{dp}{dt} = Q_1 - Q_2 + \Delta Q \tag{19}$$

$C$ : Capacity of tank($m^3/Pa$) ($C = \frac{V}{p}$  $V$ : volume of tank)

$Q_1$ : An input air flux ($m^3/s$)

$Q_2$ : An output air flux ($m^3/s$)

- An input air flux $Q_1$ is expressed as

$$Q_1 = \alpha_1 A_1 m_1 \sqrt{\frac{2(p_1 - p)}{\rho}} \tag{20}$$

Here,

$\alpha_1$ : Flow coefficient of valve

$A_1$ : pass cross-section

$m_1$ : Open degree of valve

$p_1$ : Pressure of compressed air

$p$ : Pressure of tank

$\rho$ : Density of fluids

- An output air flux $Q_2$ is expressed as

$$Q_2 = \alpha_2 A_2 m_2 \sqrt{\frac{2(p - p_2)}{\rho}} \tag{21}$$

We assume that open degree of output valve is constant.

From equation (19-21), we obtain nonlinear differential equation of tank pressure control plant.

$$\frac{dp}{dt} = \frac{1}{C}(Q_1 - Q_2 + \Delta Q) = \\
= \frac{1}{V}\alpha_1 A_1 m_1 p \sqrt{\frac{2(p_1 - p)}{\rho}} - \frac{1}{V}\alpha_2 A_2 m_2 \sqrt{\frac{2(p - p_2)}{\rho}} + \frac{1}{V} p \Delta Q \tag{22}$$

Let $x = p, u = m_1$ and $\theta = \Delta Q$, system (22) can be rewritten as follows.

$$\dot{x}(t) = f(x(t)) + g(x(t))u(t) + \omega(x(t))\theta(t) \tag{23}$$

$f(x(t)), g(x(t)), \omega(x(t))$ is as below.

$$f(x(t)) = -\frac{1}{V}\alpha_2 A_2 m_2 x \sqrt{\frac{2(x - p_2)}{\rho}} \tag{24}$$

$$g(x(t)) = \frac{1}{V}\alpha_1 A_1 x \sqrt{\frac{2(p_1 - x)}{\rho}} \tag{25}$$

$$w(x(t)) = \frac{1}{V}x \tag{26}$$

Table1. process parameter

| parameter | value | unit | description |
|---|---|---|---|
| $\alpha_1$ | 0.631811 | | Flow coefficient of input valve |
| $\alpha_2$ | 0.631811 | | Flow coefficient of output valve |
| $A_1$ | $1.9625 \times 10^{-3}$ | $m^2$ | Pass cross-section of input valve |
| $A_2$ | $1.9625 \times 10^{-3}$ | $m^2$ | Pass cross-section of output valve |
| $p_1$ | $2 \times 10^5$ | $Pa$ | Pressure of compressed air |
| $\rho$ | 3.49772 | $kg/m^3$ | Air density |
| $V$ | 2 | $m^3$ | Tank capacity |

-controller design

$$h(x) = \begin{cases} -\dfrac{L_f V + \sqrt{(L_f V)^2 + \|L_g V\|^4}}{\|L_g V\|^2}, & L_g V \neq 0 \\ 0, & L_g V = 0 \end{cases} \tag{27}$$

$$L_g V = \frac{\partial V}{\partial x} g(x), \quad L_f V = \frac{\partial V}{\partial x} f(x)$$

Where,

$$L_g V = \frac{2}{V} \alpha_1 A_1 x^2 \sqrt{\frac{2(P_1 - x)}{\rho}} \tag{28}$$

$$L_f V = -\frac{2}{V} \alpha_2 A_2 m_2 x^2 \sqrt{\frac{2(x - P_2)}{\rho}} \tag{29}$$

Fig4 shows simulation result.

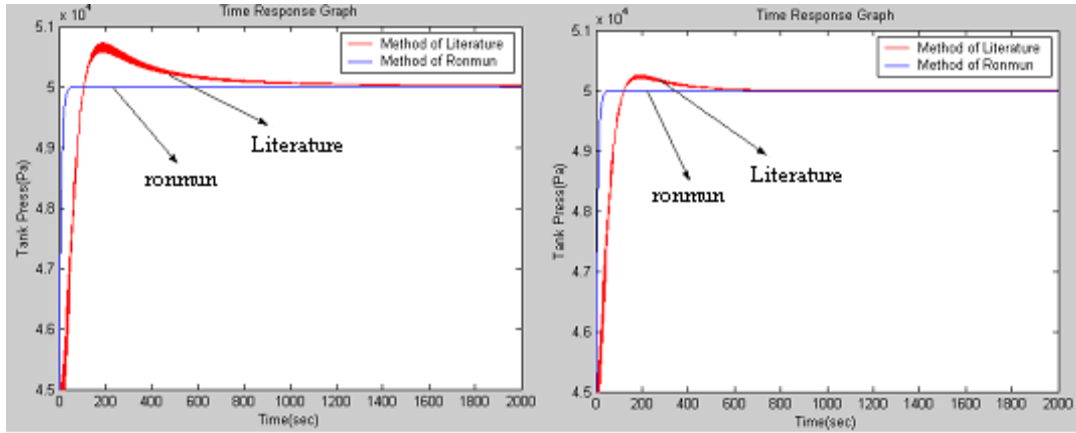

Fig4. simulation result.

$\Delta Q = 0.175, \Delta Q = 0.185$, sample time is 2s.

In order to evaluate the performance of closed system, we used costs as following.

$$J = \sum_{i=0}^{M} \left[ x(t_i)^T Q_C x(t_i) + u(t_i)^T R_C u(t_i) \right] \tag{30}$$

Where $Q_c = 1, R_c = 10^6, M = 1h$.

Table2 shows performance costs along the closed loop trajectories.

Table 2. performance costs along the closed loop trajectories.

| simu | Literature[1] | Proposed method |
|---|---|---|
| 1 | $0.1667 \times 10^{11}$ | $0.0201 \times 10^{11}$ |
| 2 | $0.3062 \times 10^{11}$ | $0.0132 \times 10^{11}$ |
| 3 | $0.6765 \times 10^{11}$ | $0.0391 \times 10^{11}$ |
| 4 | $0.9034 \times 10^{10}$ | $0.3021 \times 10^{11}$ |
| 5 | $0.4562 \times 10^{11}$ | $0.0732 \times 10^{11}$ |
| 6 | $0.3865 \times 10^{11}$ | $0.0839 \times 10^{11}$ |
| 7 | $0.4514 \times 10^{11}$ | $0.0563 \times 10^{11}$ |
| 8 | $0.4107 \times 10^{11}$ | $0.0125 \times 10^{11}$ |
| 9 | $0.9256 \times 10^{10}$ | $0.6532 \times 10^{12}$ |
| 10 | $0.7030 \times 10^{11}$ | $0.1007 \times 10^{11}$ |

**Conclusion**

In this paper, non-linear state feedback controller for input-affine nonlinear system designed taking data losses into account of control system. Data losses is compensated by means of state prediction based on model of plant, it guarantee quality of control and stability of closed system.

Simulations show that the proposed approach is superior to previous approach.